\documentclass[runningheads]{llncs}
\usepackage[title]{appendix}
\usepackage{graphicx}
\usepackage{tabularray}
\usepackage{afterpage}

\usepackage{smartdiagram}
\usesmartdiagramlibrary{additions}
\usepackage{tikz}

\usepackage{pgfplots}
\pgfplotsset{compat=1.18}
\usepgfplotslibrary{statistics}

\usepackage{subcaption}
\usepackage{caption}

\usepackage{xcolor}
\definecolor{ou-red}{cmyk}{0,0.89,0.85,0.11}
\definecolor{ou-red-darker}{cmyk}{0,1,0.96,0.29}

\usepackage{amsmath}
\usepackage{amssymb}
\usepackage{csquotes}
\usepackage{mathtools}

\usepackage{enumitem}

\usepackage{hyperref}

\usepackage{caption}

\newcommand{\appendixMeasures}{\hyperref[sect:appendix-measures]{A}}

\newcommand{\EXCLUDE}[1]{}
\newcommand{\EXCLUDEBNAICSUBMISSION}[1]{}
\newcommand{\newchunk}[1]{\vspace*{1em}\noindent\textbf{#1}\noindent}

\AtBeginDocument{
\addtolength{\abovedisplayskip}{-1ex}
\addtolength{\abovedisplayshortskip}{-1ex}
\addtolength{\belowdisplayskip}{-1ex}
\addtolength{\belowdisplayshortskip}{-1ex}
}

\begin{document}

\title{
Effective Diversification of Multi-Carousel\\ Book Recommendation}

\author{Dani\"{el} Wilten\inst{1,2}\orcidID{0009-0000-8690-0250} \and \\ Gideon Maillette de Buy Wenniger \inst{1,3}\orcidID{0000-0001-8427-7055} \and \\
Arjen Hommersom \inst{1}\orcidID{0000-0003-0125-1680}
\and \\ Paul Lucassen \inst{2}
\and \\ Emiel Poortman \inst{2}
}

\authorrunning{D. Wilten et al.}
\institute{Open University of the Netherlands, Heerlen, The Netherlands \\
\email{gemdbw AT gmail.com}\\
\email{arjen.hommersom AT ou.nl}\\
\and
OCLC, Eelde, The Netherlands \\
\email{d\_wilten AT hotmail.com}\\
\and
University of Groningen, Groningen, The Netherlands \\ 
}
%
% https://tex.stackexchange.com/questions/299813/removing-and-in-the-author-list
\renewcommand{\lastandname}{\unskip,}

\maketitle              

\begin{abstract}
    Using multiple carousels, lists that wrap around and can be scrolled, is the basis for offering content in most contemporary movie streaming platforms. Carousels allow for highlighting different aspects of users' taste, that fall in categories such as genres and authors.  
    However, while carousels offer structure and greater ease of navigation, they alone do not increase diversity in recommendations, while this is essential to keep users engaged. 
    In this work we propose several approaches to effectively increase item diversity within the domain of book recommendations, on top of a collaborative filtering algorithm. These approaches are intended to improve book recommendations in the web catalogs of public libraries.
    Furthermore, we introduce metrics to evaluate the resulting strategies, and show that the proposed system finds a suitable balance between accuracy and beyond-accuracy aspects.
    \vspace{-0.2cm}
    % \ and Recommendation systems.}
\end{abstract}

\section{Introduction}
Two-dimensional presentation of personalized recommendations allows for a more structured presentation as compared to a one-dimensional list. Prominent content platforms that offer streaming services for audio and video, including Netflix, Amazon Prime Video and HBO Max are using this two-dimensional, or multiple-carrousel, approach to recommendation. 
Correspondingly, in the e-book sector various vendors also have taken a two-dimensional approach for presenting their collection to customers.\footnote{We found Kobo, Amazon Kindle and the Barnes \& Noble NOOK platform as some examples of vendors taking this approach.}  

The commercial interest in and adoption  of two-dimensional recommendation is mirrored by an increasing interest in academic research.  Work by Rahdari et al.~\cite{Rahdari-2022} provides both a formal analysis and data-driven evaluation to explain the success of two-dimensional recommendation.  Ferrari Dacrema et al.~\cite{Ferrari-2022} add to this discussion by delving into additional factors that must be considered in the effective evaluation of two-dimensional recommendation systems, such as diversity, serendipity and novelty, also referred to as \emph{beyond-accuracy aspects}.

In the domain of public libraries, however, two-dimensional recommendation has not been adapted or explored much, despite its potential for promoting library visits and book reading. One challenge is the size of the collections relative to the number of checkouts, which is typically much larger in the domain of books and public libraries than it is in the domain of movies and streaming platforms. Moreover, there is no explicit feedback in terms of user item ratings or otherwise; there is just the implicit feedback provided by item checkouts.  To perform recommendation evaluation in absence of explicit user feedback,  from the set of user borrowed items,  a set of ground-truth `recommendation items' can be obtained using a held-out approach. Naively, system recommended items can then be compared against this ground-truth set for `correctness'. However, since the number of collection items is much larger than the thus composed ground-truth set, the chance of exactly recommending items from the ground-truth becomes very small. 

The sparsity problem is related to earlier work on dealing with label sparsity in an explicit-feedback collaborative filtering setting~\cite{bobadillaEtAlSparsityCollaborativeFiltering}. Similarly, evaluation approaches that focus on exact ground-truth matches are doomed to fail in the scenario with implicit feedback. Instead, more permissible metrics that reward recommendation of items that are similar to those in the ground-truth are required, and are proposed in this work. These solutions build upon the replacement of the single 'one-hot' item representation with a vector representation that describes items as a combination of several meaningful attributes. This conversion makes it possible to meaningfully interpret similar items that do not exactly match items from the ground truth, analogous to the way word-embeddings allow strongly related but not identical words to be identified as such~\cite{mikolov2013efficientestimationwordrepresentations}.

This paper makes several contributions. First, this research describes a prototype of a two-dimensional recommendation system for public libraries. Second, we propose and compare several strategies to effectively increase the beyond-accuracy aspects of recommendations, while retaining recommendation relevance from the user perspective. A third research contribution of this work is to provide evaluation metrics that can deal with the large collections relative to the number of user-borrowed items, while lacking explicit feedback. 

The rest of the paper is organized as follows. In Section~\ref{sect:related-work} we discuss related work, including various notions of beyond-accuracy aspects. In Section~\ref{sect:methods} we discuss the methods used to construct and evaluate multiple-carousel recommendation systems including evaluation of and improvement for beyond-accuracy aspects. Section~\ref{sect:results} presents the results, and Section~\ref{sect:discussion} provides a brief discussion and conclusion.
\section{Related Work}\label{sect:related-work}
This section describes research related to three important aspects of this research. First we consider other research related to the recommender algorithms in a public library setting. Next, research related to two-dimensional recommender systems is covered. Finally, we conclude with the beyond-accuracy metrics diversity, serendipity and novelty that help avoid (or dampen) the filter bubble effect for users.

\newchunk{Algorithm selection}\label{subsect:related-algorithm-selection}
Our work makes contributions to the research field of multiple-carrousel based (two-dimensional) recommendation. To this end, it builds upon existing algorithms for single-list (one-dimensional) recommendation. As such, a selection of adequate one-dimensional recommendation algorithms is important. 

Two works were of main importance in informing an adequate selection of recommender algorithms to use as a foundation. The %article
work by Isinkaye et al. (2015) \cite{Isinkaye-2015} provides an inventory of the basics of the recommender system domain, including the lifecycle of recommender systems and a classification of the algorithm types. The 
work by  Saytal (2019) \cite{Satyal-2019} experiments with data from the ILS used by the Vantaa City Library in Finland. The latter concludes that SVD++ (an enhanced version of Singular Value Decomposition) outperforms the other algorithms but comes at the expense of being a more resource-heavy algorithm than the other algorithms. 
Whereas Satyal's work makes assumptions on the data that is used to transform implicit feedback into explicit feedback, in contrast, our research is purely based on implicit feedback.

\newchunk{Generating two-dimensional recommendations}\label{subsect:related-two-dimensional-recommendations}
Rahdari et al. (2022) \cite{Rahdari-2022} in their article intend to \enquote{demonstrate and measure the benefits offered by the carousel-based organization of recommendations}. The most significant drawback of a single ranked list is its attempt to be \enquote{perfect} while the intent of the user is highly uncertain \cite{Rahdari-2022}. Two complementary approaches are taken to compare carousels with ranked lists: 1) Using a proposed  mathematical \textit{carousel interaction model}, 2) Conducting a series of experiments using real user data and comparing the interactions empirically.
The authors conclude that clear advantages exist of the two-dimensional carousel approach over a ranked list with respect to the criteria of \textit{navigation effort} and \textit{exiting probability}. Our research benefited from these findings as these provide empirical evidence of having a two-dimensional approach for getting recommendations presented. Our research differs in focus, since whereas \cite{Rahdari-2022} uncovers the reasons for the popularity of two-dimensional recommendations, our research is aimed at designing a solution that is capable of generating two-dimensional recommendations that are effective as well as more diverse. \\

The work by Ferrari Dacrema et al. (2022) \cite{Ferrari-2022} expands on the topic of finding the set of carousels that best suits the needs of a user. They propose to extend ranking metrics to the two-dimensional carousel setting, by not only considering the position of an item in a recommendation list, but also the position of that list in the list of recommendation lists. They also consider human preferences regarding which parts of a page to give more attention \cite{Chierichetti-2011} in the context of multiple-carrousel recommendation. 
Our work differs in the type of training data that was used during the research: where Ferrari Dacrema et al. had access to explicit feedback, our work is solely based on implicit feedback. Another difference is in the approach when selecting carousels and items: their research bases the ranking strategy on carousels that already contain items, while our strategy provides a client application with empty carousels, after which these carrousel are separately filled on-the-fly; an approach aimed at lightweight and on-demand carrousel selection and filling.

\newchunk{Beyond-accuracy aspects (diversity, serendipity, novelty)}\label{subsect:related-work-beyond-accuracy-aspects}
Kaminskas and Bridge (2017) \cite{Kaminskas-2017} present a literature survey of the most discussed beyond-accuracy objectives in recommender systems research: diversity, serendipity, novelty and coverage; of which the first three elements are user-related. 
Coverage is regarded at a system level that can be approached in two ways: 1) \textbf{user coverage} that measures the degree to which the system covers its users (i.e., is able to generate recommendations), 2) \textbf{item coverage} that measures the degree to which recommendations cover the set of available items.
They focus on \enquote{item coverage}, and present evidence on effective offline evaluation approaches for beyond-accuracy evaluation aspects. They provide a detailed comparison of known reranking approaches, which are effective approaches for improving on beyond-accuracy aspects. 
Our research differs from theirs by using  solely implicit feedback, whereas in their work implicit feedback is converted into explicit feedback using a standard frequency-based approach for such conversions. 

The work by  Sun et al. (2022) \cite{Sun-2022} uses matrix factorization to build a cross-domain recommender platform, where training data from multiple domains is combined to make better recommendations for each of the individual domains. Users overlap across the domains, but items do not. 
Their work presents an approach for combining data from \textit{cross-domains} systems to generate more diverse recommendations. In contrast,  our research aims to add diversification to a \textit{single-domain} data set.

The work of Nguyen et al. (2014) \cite{Nguyen-2014} 
covers research concluding that users following recommendations reduces the risk of a filter bubble. Their motivation is that users following the recommendations tend to provide more explicit feedback that helps a \textit{\enquote{top-n recommender list algorithm}} (which the authors implemented as an item-item algorithm) to be capable of providing more diverse recommendations. 
The authors observed a drop in content diversity in the recommendations for the group that ignored recommendations, which can be explained by the system learning more about the interests of those users and adjusting the recommendations more and more to those interests. 
Their work aims to study the effect of recommender systems on filter bubbles in general, while our research is aimed at avoiding and dampening a filter bubble effect. 
\section{Methods}\label{sect:methods}

In this section, we describe the data used in our experiments, the used recommendation algorithm, the initial methods used to select and fill recommendation carousels, the used evaluation measures, and the proposed methods to improve performance on beyond-accuracy aspects including diversity and related aspects.

\subsection{Recommendation algorithms}\label{subsect:algorithms}

The aim of our research was to investigate the applicability of a two-dimensional (multiple-carousel) recommendation approach to the public library domain. To keep the approach flexible, general and simple, we decided early on in the research to build the two-dimensional recommendation system upon one-dimensional recommendation lists produced by a well-established recommendation algorithm. To this end, we selected the used recommendation 
algorithm from the recommenders repository the Github source code repository “Best Practices on Recom-
mendation Systems” by Microsoft (2018) \cite{Microsoft-2018}. This repository contains 33 algorithms that can be grouped as: 1) Collaborative Filtering, 2) Content-Based Filtering and 3) Hybrid type algorithms. Apart from offering a good variety of recent algorithms, we selected this repository because part of the contained algorithms support working with implicit feedback, as required by our research. 

Based on an extensive literature study followed by a preliminary empirical study with several shortlisted algorithms, we decided to use the \textit{LightGCN} recommendation algorithm  \cite{He-2020} in this research.
This algorithm provides good performance combined with a much lower computational cost in comparison to some of the alternatives considered.

\subsection{Evaluation Measures}

In the subsection we discuss the used evaluation measures, which all work by comparing recommended items for a user and ground-truth items in the test set for that user. These test set items are borrowed by the respective users, but held out of training data. 
Recommendations produced by the system have a ranking order. During evaluation, we use different values $K$  and include only the top $K$ ranked predictions as a basis for evaluation, and do this for different values of $K$. The evaluation measures are then computed by comparing the set of included prediction items against the set of matching items from the test set that belong to the relevant user for that metric (sub-)computation.

The first type of measures we discuss, so-called  \emph{exact match measures}, check the exact correspondence between the recommended items and the ground-truth items. However, because the library collection is by nature much larger than the average number of borrowed items per user, this leads to sparsity, in which many valid recommendations for any user will necessarily not be borrowed by that user and hence will be missing in the ground-truth. This problem is reflected by very low scores for all of the exact match measures, discussed next. This motivated a different evaluation approach called \emph{similarity-based evaluation} discussed after the exact match measures.

\subsubsection{Exact match measures}

The used exact match evaluation measures are: precision@k, recall@k, mean average precision (MAP), and normalized discounted cumulative gain (nDCG). 
Note that each of these measures is influenced by the specified K value. 
Precision@k and recall@k measure the ratio of true-positive items amongst the top $K$ recommendations relative to the recommended $K$ items (precision) or total used items (recall).
Precision@k, recall@k and nDCG are specific to a single user, MAP averages over all users in the testset. Due to space limitations, we refer to Appendix \ref{appendix:evaluation-metrics-appendix} for further explanation and formulas of these measures which importantly can differ from expectations based on their names.

\subsubsection{Similarity-based evaluation using ABIS}\label{sect:rq2-abis}
To allow a better comparison of recommender model performance, we introduced a measure for average basic item-similarity (ABIS), to calculate the distance between two items. Using that measure, the distance between top-predicted items, and borrowed items included in the test set can be determined. The measure is very rudimentary, but it does bring the ability to do a content-based evaluation of the predicted items, which is lacking in the exact match measures. 

\paragraph{Basic Item Similarity}
The rationale behind using this similarity score is that a recommender system that predicts \enquote{\textit{Harry Potter and the Deathly Hallows}}, while the only Harry Potter title in the test items is \enquote{\textit{Harry Potter and the Goblet of Fire}} gets a zero score on that prediction from a \textit{precision} point of view. Using a similarity-based evaluation, as the two items are highly similar, the algorithm scores better than had it predicted a cooking book on Asian cuisine.

The following item features are included in this item distance measure:

\begin{description}
	\item[Author similarity] (AuthorSim) is determined by comparing only the main author. The score is either 1 for a match, or 0 for no match.
	\item[Genre similarity] (GenreSim) is determined using the Jaccard distance (see Equation~\ref{eqn:jaccard-distance} in Appendix~\appendixMeasures{}).
	\item[Subject heading similarity] (SubjectSim) is determined using the Jaccard distance. 
	\item[Age category similarity] (AgeCatSim) is determined by comparing the coarse grained age category code. The score is either 1 for a match, or 0 for no match.
	\item[Medium type similarity] (MedTypeSim) is determined by comparing the medium type code. The score is either 1 for a match, or 0 for no match.
	\item[Fiction or non-fiction similarity] (FictSim) is determined by comparing the boolean fiction code. The score is either 1 for a match, or 0 for no match.
\end{description}

To define several concepts related to item similarity, we introduce some notation.
Set of items are denoted by $I$ and $J$. A ground truth set of items is denoted by $G$ and we denote a set of predicted items by $\hat{G}$. The top-$K$ predicted items is denoted by $\hat{G}^K$. Furthermore, given a set of users $U$, the items and predicted items for $u \in U$ is denoted by $G_u$ and  $\hat{G}_u$, respectively.

The similarity between between two items \( (i,j) \), called BIS (Basic Item Similarity) can then be determined as follows, as the weighted average of the similarity of each property:

\newcommand{\BasicItemSimilarity}{BIS}

%\scriptsize
\begin{align}\label{eqn:basic-item-similarity}
 	  \text{\BasicItemSimilarity}(i,j) = 
 	  \frac{\splitfrac{\text{AuthorSim(i,j) + ( 2 $\cdot$ GenreSim(i,j) ) + SubjectSim(i,j) +}}{(\text{ 2 $\cdot$ AgeCatSim(i,j) ) + MedTypeSim(i,j) + FictSim(i,j)}}}{8}
\end{align}
%\normalsize
The age category and genre characteristics were given a boost factor of 2 as these are the most valuable in the comparison, based on expert opinion. The boost factor for age category penalizes predicted items that are intended for a different age group. The boost factor for genre penalizes predicted items that have no genre-overlap with the derived user's reading intent.

The maximum similarity of an item \( i \in I \) to a set of items $J$ is defined as follows:

\begin{equation}\label{eqn:max-basic-item-similarity}
	\text{MaxBIS}(i, J) = \max_{j \in J} \text{\BasicItemSimilarity}(i,j)
\end{equation}
Similarly, the average similarity score of an item \( i \in I \) to a set of items $J$ is defined as follows:

\begin{equation}\label{eqn:item-to-set-average-similarity}
 	\text{AvgBIS}(i, J) = \frac{1}{|J|} \sum_{j \in J} \text{\BasicItemSimilarity}(i, j)
\end{equation}
which can be generalized to give an average similarity between two sets of items: 
\begin{equation}\label{eqn:set-to-set-similarity} 
 	\text{AvgSetBIS}(I, J) = \frac{1}{|I|} \sum_{i \in I} \text{AvgBIS}(i, J) 	
\end{equation}

Given a single set of items $I$, we define the internal similarity score 
to indicates the average pairwise similarity for all unique item pairs:

\begin{equation}\label{eqn:internal-item-set-similarity}
	\text{InternalSimilarity}(I) = \frac{1}{|I|^2} \sum_{i\in I} \sum_{j\in I} \text{\BasicItemSimilarity}(i, j)
\end{equation}

Using the the maximum similarity for an item to a set of items, we can determine the average similarity score for one user's \( K \) top-predicted items, to the set of test-items that were held back for evaluation purposes during the offline experiments:

\begin{equation}\label{eqn:basic-item-similarity-at-k}
	\text{BIS@K}(G, \hat{G}^K) = \frac{1}{K} \sum_{\hat{g} \in \hat{G}^K} \text{MaxBIS}(\hat{g}, G)
\end{equation}

This definition is used to determine a similarity-based performance score for a recommender model for a single user. In addition, we take the average score for \textit{all users} in the data set, which we refer to as the \textbf{ABIS} (Average Basic Item Similarity) score given a user set $U$.

\begin{equation}\label{eqn:average-basic-item-similarity}
	\text{ABIS}(U) = \frac{1}{|U|} \sum_{u \in U} \text{BIS@K}(G_{u}, \hat{G}^K_{u})
\end{equation}

\subsection{Two-dimensional recommendation strategies}

The LightGCN algorithm that we employ produces a simple graph convolutional network model designed for collaborative filtering. The algorithm only takes implicit feedback as input, no user or item features can be included in the training. Item features are used however in the recommender carousel filling stage, when after adding recommended items, additional items are added from the collection using various strategies aimed to improve beyond-accuracy aspects of the selection.

\subsubsection{Basic strategy for Carousel Selection and Filling}\label{sect:method-selection-strategies-basic}

We included a limited set of four different carousel types to instantiate specific carousels from. The aim was to use a small selection of fundamental types for which  sufficient data exists to inform selection choices, and which sufficiently supports content variety. This motivated use of the following list of used carousel types: Genre carousel, Subject carousel, Author carousel, and Cyclic-event carousel types.

Next, for carousel selection we used a heuristic procedure that builds upon: (1) the recommendations made for users by the trained one-dimensional recommendation system, and (2) the recently borrowed items of a user. It does so by using frequencies for the genre, subject, and author features of these items. Specifically, we leveraged the user-specific top-5 predicted items, and the top-5 most recently borrowed items, to identify fitting genres, subjects, and authors given these statistics. 

\newchunk{Users with borrowing history} For a \textit{user with (borrowing) context}  the carousel-selection procedure is as follows:
\begin{itemize}
	\item \textbf{Fetch the user's 5 top-predicted items and 5 most recently borrowed items}
	\item \textbf{Select genre-type carousels}
	\begin{itemize}
		\item From top predicted items
		\begin{itemize}
			\item \textit{Pick 3 most frequent individual genres}
			\item \textit{Pick 2 most frequent genre combinations}
		\end{itemize}
		\item From recently borrowed items (excluding already picked values)
		\begin{itemize}
			\item \textit{Pick 3 most frequent individual genres}
			\item \textit{Pick 2 most frequent genre combinations}
		\end{itemize}
	\end{itemize}
	\item \textbf{Select subject-type carousels}
	\begin{itemize}
		\item From top predicted items
		\begin{itemize}
			\item \textit{Pick 3 most frequent subjects}
		\end{itemize}
		\item From recently borrowed items (excluding already picked values)
		\begin{itemize}
			\item \textit{Pick 3 most frequent subjects}
		\end{itemize}
	\end{itemize}
	\item \textbf{Select author-type carousels}
	\begin{itemize}
		\item From top predicted items
		\begin{itemize}
			\item \textit{Pick 3 most frequent authors}
		\end{itemize}
		\item From recently borrowed items (excluding already picked values)
		\begin{itemize}
			\item \textit{Pick 3 most frequent authors}
		\end{itemize}
	\end{itemize}
\end{itemize}

\newchunk{Users without context} For users without borrowing history, a similar procedure is used, but based on different item sets to compute feature-frequency statistics. Here,to replace the non-available (no-history) user predictions, we pool the top-1 recommendations for every user with history and use this as the frequency statistics basis for the carousels based on recommendation predictions. As a replacement for recently borrowed items by the specific user, we instead use the 1000 most recently borrowed items by all users. Using those, in place of the top-5 user-predicted and top-5 recently user-borrowed items, the procedure is for the rest mostly the same as the one for users with context. The only other small difference is that in this case carousels for frequent \emph{individual} genres are omitted, with only those for genre-combinations included.

\newchunk{Item Selection}\label{chunk:item_selection_basic_strategy} After selecting specific carousels for a user, filling them was done following a simple procedure. For users with borrowing-history, up to 15 highest ranked, not previously-borrowed items matching the carousel constraint (e.g. matching author) were selected from the user-recommendation list. If necessary, these were further topped-up to 15 (or as close as possible) using matching items from the rest of the collection. For users without borrowing history, this procedure simplified to selecting as many as possible (up to 15) carousel-matching items from the entire collection. \EXCLUDEBNAICSUBMISSION{\marginpar{For non-context users, is the selection fully random, or does it still involve some statistics as in the case of the carousel selection for non-context users?}}  

\subsubsection{Improving carousel items selection for diversity}\label{sect:method-selection-strategies-diversity}
To improve diversity of selected items, the original selection strategy is changed to select a maximum of 10 instead of 15 predicted items with the highest prediction score, and then pick 100 items from the library collection matching the carousel constraints, from which a selection will be made to end up with the desired total of 15 carousel items. This selection from the 100 library collection items will be focused on \textit{diversity} with respect to the already selected carousel items.

Diversity is defined by Merriam-Webster\cite{site:merriamwebster} as \enquote{\textit{the condition of having or being composed of differing elements : variety}} . The goal is then to achieve diverse carousels by achieving \enquote{variety} among the carousel items. In the original strategy, no attention to item features is paid beyond the fact that the items need to match the carousel constraints. In this diversity-focused strategy, we selected the additional items from the 100 library collection items that are most dissimilar to the items already selected.
This is implemented by using the average similarity defined in Equation~\ref{eqn:item-to-set-average-similarity}, by greedily adding items from the 100 selected items to the itemlist that has the lowest average similarity score to the itemlist until 15 items are selected.

\subsubsection{Improving carousel items selection for serendipity}\label{sect:method-selection-strategies-serendipity}
Serendipity is typically described as: \enquote{\textit{finding something valuable or agreeable that was not sought for}}. It is the event when a user is recommended a relevant item that he finds surprising. In their work, Kaminskas and Bridge (2017, pg.23) \cite{Kaminskas-2017} state: \enquote{\textit{measuring surprise is based on the intuition that a recommendation is surprising if it is unlike any item the user has seen before.}}. Using that statement as the starting point for achieving serendipity, we concluded that serendipity could be achieved by including recommended items that are most dissimilar to the items that the user has borrowed. By introducing a variant of the diversity strategy defined in Section~\ref{sect:method-selection-strategies-diversity} that selects items that are most dissimilar to the transactional history of the user, we aimed at including serendipitous items in the carousels.

In the prototype software, support for a new item selection strategy that focuses on serendipity was added. The strategy is similar to the diversity strategy for the most part, but instead of finding 5 additional items among a 1,000 item set from the library collection that are most \textit{dissimilar to the already selected carousel items}, the new strategy looks for 5 additional items that are most \textit{dissimilar to the borrowing history of the user}.

\subsubsection{Improving carousel items selection for novelty}\label{sect:method-selection-strategies-novelty}
Novelty is typically explained as finding something new or unusual. For this step we focused on finding something \textit{new}, as we believed finding something \textit{unusual} is already covered through the efforts to include serendipitous items. We can look for items that are assumed to be new to the user in three ways:
\begin{enumerate}
	\item By selecting recently added items from the library collection.
	\item By selecting recently first-time published items from the library collection.
	\item By selecting long-tail items from the library collection (i.e., items that are not getting borrowed in recent years).
\end{enumerate}

Recently first-time published items are a subset of recently added items, since a recently first-time published item can only have been recently added, but recently added items may include reprints of items that were originally published many years ago. Combining both these types in a single strategy is expected to be the most effective approach, since it mitigates the issue of not having 
sufficient recently first-time published items in the collection matching the carousel constraints.
We limit the scope for novelty to \textit{all recently added items, where recently first-time published items are prioritized over recently added items that were published longer in the past}. Including long-tail items is left for future research. 

\subsubsection{Combining the carousel items selection strategies}\label{sect:method-selection-strategies-combined}
In their work,  Kaminskas and Bridge (2017) \cite{Kaminskas-2017} state that beyond accuracy objectives can be in conflict. Selecting items recently added to the collection to improve the novelty score could dampen the diversity of the carousel. In this section, the three item selection strategies from the previous sections are combined in a new item selection strategy that should deliver a well-balanced result.

The combined strategy selects 15 items for a carousel, given that sufficient items exist that match the constraints of the carousel. Instead of selecting up to 15 items from the top-predicted items for the user, the combined strategy will only select 9 top-predicted items, and will then loop through the novelty strategy, serendipity strategy and diversity strategy to add 1 item per strategy until 15 items have been reached or no additional items exist for adding.

\subsection{Experimental Setup}
\paragraph{Data selection}\label{sect:rq1-data-selection}
Experiments are conducted to obtain performance results for different strategies. 
For the experiments, we use a (real-world) public library data set that contains: 1) \textbf{Transactional data} containing a user-ID, title-ID and a checkout timestamp for checkout transactions.
2) \textbf{Item data} containing the title, author, year of publication, and genres for titles included in the transactions.
 
For computational reasons, a subset of the total available transactions is used in our experiments. First, more recent transactions are prioritized, to better model user's current interest.
Second, transactions of users with a very high or very low number of annual loans are excluded. We allowed a maximum of 50 annual loans per user, to filter out non-individual user transactions (i.e. organizations). We also required a minimum of 10 annual loans to filter out user transactions which add little informative value when training a recommender system. This resulted in a dataset consisting of 125,069 transactions from 5,539 users that contains 52,596 items, which we refer to as the \emph{Delta} dataset.

Besides the Delta dataset, we made use of the public Movielens-100k dataset which holds 100,836 transactions. Both datasets have some different characteristics. The movielens database contains 100,836 ratings across 9,742 movies by 610 users. The data in the six public library data sets is much more sparse: on average, each Movielens user is included in 0.1639\% of the transactions and each Movielens item in 0.0103\% of the transactions, whereas each library user is included in 0.0024\% of the transactions and each library item is included in 0.0002\% of the transactions.

\paragraph{Training and Testing Sets}
For our experiments, the data split in a part used for training and a part used for testing. 
Out of the 125,069 transactions in the delta dataset, we created a test set consisting of 100 users with on average 5 transactions per user, for a total of 500 test transactions. The remaining 124,569 were used for training the recommendation model.

\paragraph{Research parameters}\label{sect:rq3-research-parameters}
Within the experiments, decisions on various parameters were made. For filling carousels, we selected 
a total of 15 items. For the specialized beyond-accuracy strategies, 10 items were selected from the top matching recommendations and the remaining ones selected with the strategy. For the combined strategy, 9 items were selected based on the recommendations and the remaining 6 with the strategy. The serendipity strategy and evaluation considered a maximum of 200 most recent transactions. All strategies were evaluated by averaging the relevant metrics over a test set of 100 users. The experiments were conducted on a laptop running Microsoft Windows with i7 processor and 32GB of RAM.
\begingroup
\scriptsize
\begin{longtblr}[
  caption = {Table with precision@k, recall@k, MAP and nDCG scores (best scores in boldface)},
  label = {tab:precision-recall-map-ndcg-scores},
]{
   colspec = {X[0.22,j] X[0.10,j] X[0.15,j] X[0.15,j] X[0.14,j] X[0.14,j]},
  rowhead = 1,
  hlines,
  row{even} = {red9},
  row{1} = {ou-red},
}

 \textcolor{white}{Data set} & \textcolor{white}{K} & \textcolor{white}{Precision@k} & \textcolor{white}{Recall@k} & \textcolor{white}{MAP} & \textcolor{white}{nDCG} & \textcolor{white}{Runtime}\\
\hline
Delta & 10 & \textbf{0.019101} & 0.035401 & 0.014583 & 0.030948 & 4min\\ 
Delta & 25 & 0.013418 & 0.061627 & 0.016934 & 0.041613 & 4min\\ 
Delta & 50 & 0.009832 & 0.090055 & 0.018203 & 0.050984 & 5min\\ 
Delta & 100 & 0.006907 & 0.126388 & 0.019104 & 0.061058 & 5min\\ 
Delta & 250 & 0.004192 & \textbf{0.192561} & \textbf{0.019943} & 0.076426 & 5min\\ 
\textit{Movielens-100k} & \textit{10} & \textit{0.360976} & \textit{0.196052} & \textit{0.121633} & \textit{0.417629} & \textit{n/a}\\
\end{longtblr}
\endgroup

\section{Results}\label{sect:results}

\subsection{Results for one-dimensional recommendation system evaluation}
Using the results of the offline experiments, in this section the performance of the trained recommender models is evaluated using the earlier discussed evaluation measures. 

\paragraph{Exact match measures scores}
Table \ref{tab:precision-recall-map-ndcg-scores} shows the results for the evaluation of LightGCN using the exact match evaluation measures. As mentioned before in the methods section, for all settings the scores for the exact match measures are very low. This motivated the use of the ABIS scores, which by working with item similarity rather than exact measures, aims to solve aforementioned sparsity problems during the evaluation. 

\paragraph{ABIS scores}
Using Equation~\ref{eqn:average-basic-item-similarity}, the ABIS score for LightGCN algorithm (with different values for $K$) for which top-predicted items have been stored in the local database were calculated. As a baseline, we randomly chose items from the data set and assigned a gradually descending prediction score starting at 1, and then declining by 0.001 values per item. We took this approach for different $K$ values $(10, 25, 50, 100, 250)$, and used the LightGCN transactional test data to calculate the ABIS scores for this algorithm, see the right two columns in Table~\ref{tab:abis-scores}.

\afterpage{
\begin{center}
\begin{minipage}{.5\textwidth}
\begingroup
\scriptsize
\begin{longtblr}[
  caption = {Table with ABIS scores},
  label = {tab:abis-scores},
]{
colspec = {X[0.10,j] || X[0.35,j] X[0.20,r] || X[0.35,r] X[0.20,j]},
  rowhead = 1,
  hlines,
  row{even} = {red9},
  row{1} = {ou-red},
}
\textcolor{white}{K} & \textcolor{white}{Algorithm}  &  \textcolor{white}{ABIS} & \textcolor{white}{Algorithm}  &  \textcolor{white}{ABIS}\\
\hline
10 & LightGCN &  0.699  & Random  &  0.691 \\ 
25 & LightGCN &  0.724 & Random  &  0.662 \\ 
50 & LightGCN &  0.739  & Random  &  0.679 \\ 
100 & LightGCN & 0.752  & Random  &  0.691 \\
250 & LightGCN &  0.766 & Random  &  0.710 \\
\end{longtblr}
\endgroup
\end{minipage}%
\begin{minipage}{0.48\textwidth}
\begin{center}
 \scalebox{0.9}{
\begin{tikzpicture}
    \begin{axis}[
        width=5cm,           % Reduce the width of the plot
        height=4.5cm,          % Reduce the height of the plot
        boxplot/draw direction=x,
        xlabel={Internal similarity},
        ylabel={Carousel items selection strategy},
        ytick={1, 2, 3, 4, 5},
        yticklabels={Original, Combined, Diversity, Serendipity, Novelty},
        xmin=0, xmax=1.0,
        xtick={0.0, 0.2, 0.4, 0.6, 0.8, 1.0},
        xticklabel style={/pgf/number format/fixed, /pgf/number format/precision=3},
        grid=both,
        boxplot/box extend=0.3
    ]
        % Data for Original strategy
        \addplot+[
            boxplot prepared={
                median=0.501,
                upper quartile=0.561,
                lower quartile=0.436,
                upper whisker=0.844,
                lower whisker=0.151
            },
            draw=blue,
            fill=blue!30
        ] coordinates {};

        % Data for Combined strategy
        \addplot+[
            boxplot prepared={
                median=0.379,
                upper quartile=0.465,
                lower quartile=0.316,
                upper whisker=0.844,
                lower whisker=0.114
            },
            draw=green,
            fill=green!30
        ] coordinates {};

        % Data for Diversity strategy
        \addplot+[
            boxplot prepared={
                median=0.377,
                upper quartile=0.472,
                lower quartile=0.301,
                upper whisker=0.844,
                lower whisker=0.094
            },
            draw=red,
            fill=red!30
        ] coordinates {};

        % Data for Serendipity strategy
        \addplot+[
            boxplot prepared={
                median=0.431,
                upper quartile=0.505,
                lower quartile=0.347,
                upper whisker=0.844,
                lower whisker=0.127
            },
            draw=orange,
            fill=orange!30
        ] coordinates {};

        % Data for Novelty strategy
        \addplot+[
            boxplot prepared={
                median=0.453,
                upper quartile=0.523,
                lower quartile=0.396,
                upper whisker=0.844,
                lower whisker=0.151
            },
            draw=cyan,
            fill=cyan!30
        ] coordinates {};
    \end{axis}
\end{tikzpicture}
}
\captionsetup{justification=centering}

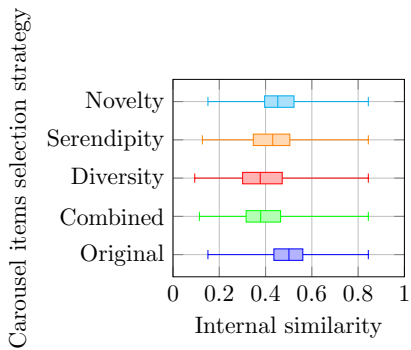
\captionof{figure}{Internal carousel similarity for 100 users}
\label{fig:rq3-carousel-internal-similarity-effect-all-strategies-100-users}
\end{center}
\end{minipage}
\end{center}
}

We conclude from these results that the items recommended by a model trained using an algorithm are on average more similar to the test items for all users, than the randomly predicted items. 
\subsection{Comparison of Two-Dimensional Recommendation Strategies}\label{sect:rq3-results}

This section describes the comparison of different recommendation strategies, in particular for  improving the item selection strategy on beyond-accuracy aspects, i.e., \textit{diversity}, \textit{serendipity} and \textit{novelty}.  For all carousels, the requirement for any item is that the item must meet the constraints for that carousel. This requirement was maintained while aiming for increased beyond-accuracy aspects, such as diversity: returning an item written by \textit{J. J. R. Tolkien} in a \enquote{author-type carousel} for \textit{Marian Keyes} may increase diversity, but the item is not applicable for the given carousel.

\subsubsection{Impact of item selection strategies on diversity}\label{sect:rq3-results-diversity}

In Fig.~\ref{fig:rq3-carousel-internal-similarity-effect-all-strategies-100-users}, the internal carousel similarity is visualized, measured for 100 users over a total of 1.614 carousels, with each carousel containing up to 15 items. A lower similarity score indicates a higher diversity, which shows that the original item selection strategy from Section \ref{sect:method-selection-strategies-basic} is outperformed on this aspect by all other strategies. For diversity, the two best performing strategies are the \textit{combined} strategy and the \textit{diversity} strategy.

One thing that stands out is that the strategy specialized on diversity is slightly outperformed by the combined strategy. In the boxplot, the upper quarter for the diversity strategy is at 0.472, while the upper quarter for the combined strategy is at 0.465. 
This should not be overinterpreted however, as it is likely an artefact of the fact that for the combined strategy we selected 9 items based on recommendations, adding 6 based on the strategy, whereas for the diversity strategy we selected 10 items based on the recommendations adding only 5 based on the strategy.

When comparing the results of the original strategy against those for the diversity strategy, a significant reduction in carousel-items similarity within carousels is shown. The median for InternalSimilarity for the diversity strategy is over 18 percent lower than for the original strategy. These numbers were determined by analyzing 1.348 carousels generated for 100 users.

% Fix to get the figure number right, and not skip over number 2
\addtocounter{figure}{-1}%
\begin{figure*}[t!]
\begin{minipage}{0.48\textwidth}
\begin{subfigure}[b]{\textwidth} 
\scalebox{0.9}{
\begin{tikzpicture}
    \begin{axis}[
        width=5cm,           % Reduce the width of the plot
        height=2.5cm,          % Reduce the height of the plot
        boxplot/draw direction=x,
        xlabel={Similarity},
        ylabel={Carousel-items strategy},
        ylabel style={font=\small},
        ytick={1, 2},
        yticklabels={Original, Diversity},
        yticklabel style={font=\small}, % Change font size here
        xmin=0, xmax=1.0,
        xtick={0.0, 0.2, 0.4, 0.6, 0.8, 1.0},
        xticklabel style={/pgf/number format/fixed, /pgf/number format/precision=3},
        grid=both,
        boxplot/box extend=0.3,
        ylabel style={text width=1.5cm,align=left}
    ]
        % Data for Original strategy
        \addplot+[
            boxplot prepared={
                median=0.502,
                upper quartile=0.548,
                lower quartile=0.448,
                upper whisker=0.703,
                lower whisker=0.287
            },
            draw=blue,
            fill=blue!30
        ] coordinates {};

        % Data for Diversity strategy
        \addplot+[
            boxplot prepared={
                median=0.371,
                upper quartile=0.438,
                lower quartile=0.213,
                upper whisker=0.625,
                lower whisker=0.240
            },
            draw=red,
            fill=red!30
        ] coordinates {};
    \end{axis}
\end{tikzpicture}
}
\caption{Genre carousel type}
\end{subfigure}\\
\begin{subfigure}[b]{\textwidth}  
\scalebox{0.9}{
\begin{tikzpicture}
    \begin{axis}[
        width=5cm,           % Reduce the width of the plot
        height=2.5cm,          % Reduce the height of the plot
        boxplot/draw direction=x,
        xlabel={Similarity},
        ylabel={Carousel-items strategy},
        ylabel style={font=\small},
        ytick={1, 2},
        yticklabels={Original, Diversity},
        yticklabel style={font=\small}, % Change font size here
        xmin=0, xmax=1.0,
        xtick={0.0, 0.2, 0.4, 0.6, 0.8, 1.0},
        xticklabel style={/pgf/number format/fixed, /pgf/number format/precision=3},
        grid=both,
        boxplot/box extend=0.3,
        ylabel style={text width=1.5cm,align=left}
    ]
        % Data for Original strategy
        \addplot+[
            boxplot prepared={
                median=0.456,
                upper quartile=0.513,
                lower quartile=0.388,
                upper whisker=0.767,
                lower whisker=0.151
            },
            draw=blue,
            fill=blue!30
        ] coordinates {};

        % Data for Diversity strategy
        \addplot+[
            boxplot prepared={
                median=0.297,
                upper quartile=0.383,
                lower quartile=0.230,
                upper whisker=0.766,
                lower whisker=0.094
            },
            draw=red,
            fill=red!30
        ] coordinates {};
    \end{axis}
\end{tikzpicture}
}
\caption{Subject carousel type}
\end{subfigure} \\

\begin{subfigure}[b]{\textwidth}  
\scalebox{0.9}{
\begin{tikzpicture}
    \begin{axis}[
        width=5cm,           % Reduce the width of the plot
        height=2.5cm,          % Reduce the height of the plot
        boxplot/draw direction=x,
        xlabel={Similarity},
        ylabel={Carousel-items strategy},
        ylabel style={font=\small},
        ytick={1, 2},
        yticklabels={Original, Diversity},
        yticklabel style={font=\small}, % Change font size here
        xmin=0, xmax=1.0,
        xtick={0.0, 0.2, 0.4, 0.6, 0.8, 1.0},
        xticklabel style={/pgf/number format/fixed, /pgf/number format/precision=3},
        grid=both,
        boxplot/box extend=0.3,
        ylabel style={text width=1.5cm,align=left}
    ]
        % Data for Original strategy
        \addplot+[
            boxplot prepared={
                median=0.575,
                upper quartile=0.630,
                lower quartile=0.516,
                upper whisker=0.844,
                lower whisker=0.188
            },
            draw=blue,
            fill=blue!30
        ] coordinates {};

        % Data for Diversity strategy
        \addplot+[
            boxplot prepared={
                median=0.495,
                upper quartile=0.585,
                lower quartile=0.428,
                upper whisker=0.844,
                lower whisker=0.188
            },
            draw=red,
            fill=red!30
        ] coordinates {};
    \end{axis}
\end{tikzpicture}
}
\caption{Author carousel type}
\end{subfigure}
\captionsetup{justification=centering}
\captionof{figure}{Item similarity strategies effect measured per carousel type}
\label{fig:rq3-effect-diversity-strategy-per-carousel-type}
%\end{figure}
\end{minipage}
\hspace{0.5cm}
\begin{minipage}{0.48\textwidth}
%\begin{figure}[h!tbp]
\begin{center}
\scalebox{0.9}{
\begin{tikzpicture}
    \begin{axis}[
        width=5cm,           % Reduce the width of the plot
        height=4.5cm,          % Reduce the height of the plot
        boxplot/draw direction=x,
        xlabel={Transactions similarity},
        ylabel={Carousel items selection strategy},
        ytick={1, 2, 3, 4, 5},
        yticklabels={Original, Combined, Diversity, Serendipity, Novelty},
        xmin=0, xmax=1.0,
        xtick={0.0, 0.2, 0.4, 0.6, 0.8, 1.0},
        xticklabel style={/pgf/number format/fixed, /pgf/number format/precision=3},
        grid=both,
        boxplot/box extend=0.3,
        ylabel style={text width=2.5cm,align=right}
    ]
        % Data for Original strategy
        \addplot+[
            boxplot prepared={
                median=0.311,
                upper quartile=0.378,
                lower quartile=0.218,
                upper whisker=0.653,
                lower whisker=0.020
            },
            draw=blue,
            fill=blue!30
        ] coordinates {};

        % Data for Combined strategy
        \addplot+[
            boxplot prepared={
                median=0.258,
                upper quartile=0.326,
                lower quartile=0.182,
                upper whisker=0.653,
                lower whisker=0.023
            },
            draw=green,
            fill=green!30
        ] coordinates {};

        % Data for Diversity strategy
        \addplot+[
            boxplot prepared={
                median=0.253,
                upper quartile=0.324,
                lower quartile=0.183,
                upper whisker=0.653,
                lower whisker=0.029
            },
            draw=red,
            fill=red!30
        ] coordinates {};

        % Data for Serendipity strategy
        \addplot+[
            boxplot prepared={
                median=0.241,
                upper quartile=0.318,
                lower quartile=0.156,
                upper whisker=0.653,
                lower whisker=0.004
            },
            draw=orange,
            fill=orange!30
        ] coordinates {};

        % Data for Novelty strategy
        \addplot+[
            boxplot prepared={
                median=0.295,
                upper quartile=0.364,
                lower quartile=0.209,
                upper whisker=0.653,
                lower whisker=0.006
            },
            draw=cyan,
            fill=cyan!30
        ] coordinates {};
    \end{axis}
\end{tikzpicture}
}
\captionsetup{justification=centering}
\caption{Carousel similarity with transaction items for 100 users, based on Equation~\ref{eqn:set-to-set-similarity}}
\label{fig:rq3-carousel-transactions-similarity-effect-all-strategies-100-users}
\end{center}
%%%%%%%%%%%%%%%%%%%
\begin{center}
\scalebox{0.9}{
\begin{tikzpicture}
    \begin{axis}[
        %width=5.4cm,           % Reduce the width of the plot
        %height=5.5cm,          % Reduce the height of the plot
        width=5cm,           % Reduce the width of the plot
        height=4.5cm,          % Reduce the height of the plot
        boxplot/draw direction=x,
        xlabel={\% Novelty items},
        ylabel={Carousel items selection strategy},
        ytick={1, 2, 3, 4, 5},
        yticklabels={Original, Combined, Diversity, Serendipity, Novelty},
        xmin=0, xmax=100,
        xtick={0, 20, 40, 60, 80, 100},
        xticklabel style={/pgf/number format/fixed, /pgf/number format/precision=1},
        grid=both,
        boxplot/box extend=0.3,
        ylabel style={text width=2.5cm,align=right}
    ]
        % Data for Original strategy
        \addplot+[
            boxplot prepared={
                median=0.0,
                upper quartile=6.7,
                lower quartile=0.0,
                upper whisker=100.0,
                lower whisker=0.0
            },
            draw=blue,
            fill=blue!30
        ] coordinates {};

        % Data for Combined strategy
        \addplot+[
            boxplot prepared={
                median=20.0,
                upper quartile=40.0,
                lower quartile=6.7,
                upper whisker=100.0,
                lower whisker=0.0
            },
            draw=green,
            fill=green!30
        ] coordinates {};

        % Data for Diversity strategy
        \addplot+[
            boxplot prepared={
                median=0.0,
                upper quartile=6.7,
                lower quartile=0.0,
                upper whisker=100.0,
                lower whisker=0.0
            },
            draw=red,
            fill=red!30
        ] coordinates {};

        % Data for Serendipity strategy
        \addplot+[
            boxplot prepared={
                median=0.0,
                upper quartile=6.7,
                lower quartile=0.0,
                upper whisker=100.0,
                lower whisker=0.0
            },
            draw=orange,
            fill=orange!30
        ] coordinates {};

        % Data for Novelty strategy
        \addplot+[
            boxplot prepared={
                median=40.0,
                upper quartile=53.3,
                lower quartile=6.7,
                upper whisker=100.0,
                lower whisker=0.0
            },
            draw=cyan,
            fill=cyan!30
        ] coordinates {};
    \end{axis}
\end{tikzpicture}
}
\captionsetup{justification=centering}
\captionof{figure}{Percentage of novelty items in carousels for 100 users}
\label{fig:rq3-carousel-novelty-effect-all-strategies-100-users}
\end{center} 
\end{minipage}
\end{figure*}

To check whether all of the carousel types were impacted significantly, or if it may not be worth the additional processing effort to optimize for diversity for some of the carousel types, the measurement was determined per carousel type for both strategies in Fig.~\ref{fig:rq3-effect-diversity-strategy-per-carousel-type}.

What can be observed is that the impact on the median for genre-type (30 percent less similarity) and subject-type (27 percent less similarity) carousels is significant, while for author-type carousels the impact of the diversity-focused strategy is much lower (only 9 percent less similarity). 
We believe this can partly be explained by the fact that many authors do not have sufficient items in the library collection to lead to a different selection for both strategies. For author carousel where this is not the case, using the diversity strategy is futile. Another related factor likely to contribute is that the author carousel implies by nature a strong constraint, since many authors will write books that share the same values for features that influence the ABIS similarity score.

\subsection{Impact of item selection strategies on serendipity}
To measure serendipity, we compute the similarity between the selected items $I$ in a carousel with  the 200 most recent transaction items ($T^{R}$) for that user, noted $\textrm{AvgSetBIS}(I,T^{R})$, see Equation~\ref{eqn:set-to-set-similarity}. If the similarity is \textbf{lower}, the recommended items are more different to the items the user has previously borrowed, resulting in an increased chance that the recommended items are serendipitous.

In Fig.~\ref{fig:rq3-carousel-transactions-similarity-effect-all-strategies-100-users}, the similarity between the selected carousel items with the most recent 200 transaction items is visualized, measured for 100 users over a total of 1.680 carousels, with each carousel containing up to 15 items. A lower similarity score indicates a larger chance of the recommended items being serendipitous. The top three best performing strategies for serendipity are: 1) the \textit{serendipity} strategy, 2) the \textit{diversity} strategy and 3) the \textit{combined} strategy. The difference in scores between these three strategies is minimal.

\pgfplotsset{width=7cm}

\subsection{Impact of item selection strategies on novelty}
To measure novelty, we measure the percentage of novelty items among the recommended items in a carousel. The dataset used was created in July of 2023; we considered all items added to the data set in 2023 as novelty items.

In Fig.~\ref{fig:rq3-carousel-novelty-effect-all-strategies-100-users}, the percentage of novelty items in the carousels is visualized, measured for 100 users over a total of 1.680 carousels, with each carousel containing up to 15 items. The specialized \textit{novelty} strategy achieves the highest percentage novelty items in carousels, with a median at 40 \% novelty items. But with the median at 20 \% novelty items, the \textit{combined} strategy is still significantly outperforming the other strategies.
\section{Discussion and Conclusions}
\label{sect:discussion}

In this paper, we have discussed a multi-carousel book recommendation system that takes into account various beyond-accuracy aspects.  As a foundation for this system, we trained a recommendation model using the LightGCN collaborative filtering algorithm. Due to the inherent sparsity of public library data book borrowing data, in which members by nature will have borrowed only a fraction of the entire large collection, naive application of standard metrics leads to very low scores making it hard to see if the recommendation system performs properly. By instead using an item-similarity based metric, the ABIS score, we could see a modest but clear improvement over a baseline model that performs random item selection.

Building on item selection base on the trained recommendation-model, we next explored several strategies to improve beyond-accuracy aspects of the items selected to fill up recommendation carousels in a two-dimensional library book recommendation system. The results of the experiments show that the devised individual strategies are successful in improving diversity, serendipity and novelty of the added items 
individually. More strikingly, our experiments also show that a combined strategy offers a good trade-off that increases diversity, serendipity and novelty all at the same time in comparison to the original baseline approach that does not consider beyond-accuracy aspects in the item selection.

In future work, we aim to further investigate the sensitivity of various choices within the system, such as the weighting factors in the BIS score, as this may slightly influence the results. Crucially, this will also require tests which actual users to find out which strategy in the end will be preferred. 
Furthermore, our objective is to implement the proposed system in practice. 

\bibliographystyle{splncs04}
\bibliography{report}

\newpage
% See: https://tex.stackexchange.com/questions/179997/using-package-appendix-with-llncs
\begin{subappendices}
\renewcommand{\thesection}{\Alph{section}}%
% or try \arabic{section}

\setcounter{chapter}{2}
\setcounter{section}{0} % Reset to the first appendix section
-\section{Evaluation and similarity measures}
\label{appendix:evaluation-metrics-appendix}
%% Source for used Latex equations: https://permetrics.readthedocs.io/en/v1.3.1/pages/regression.html
% This appendix provides an overview of the different types of measures that are relevant for creating recommender systems, and a listing of the most common available measures per measure type.
This appendix provides more details regarding the used evaluation metrics and similarity measures used in this thesis as well as some related metrics that are not used but serve as context. Measures for systems that produce ratings, while relevant for recommendation systems in general, are not discussed here since these are not used for this work.

 \subsection{Similarity measures}\label{subsect:appendix-measures-similarity}
% Similarity measures are used to measure distance between users or items.
 \begin{description}
	\item[Jaccard distance] or \textit{Jaccard index} is based on common preferences between users, which could be expressed as the item sets $x$ and $y$ they have borrowed \cite{Satyal-2019}.
	\begin{equation} \label{eqn:jaccard-distance}
 	\text{Jaccard Distance}(x,y) = \frac{|x \cap y|}{|x \cup y|}
	\end{equation}
 \end{description}
 
\subsection{Usage prediction measures}\label{subsect:appendix-measures-predicted-usage}
When the recommender system does not produce ratings but instead suggests items that may be of interest to the user, we are more interested in whether the system properly predicts items that the user will use.

In offline evaluation we can measure this by hiding items a user selected from the data set used for training the model, and then have the recommender system predict items the user would be interested in. This would lead to four possible outcomes:
\begin{itemize}
	\item True-positive when the item was recommended and used.
	\item False-positive when the item was recommended but not used.
	\item False-negative when the item was not recommended but was used.
	\item True negative when the item was not recommended and not used.
\end{itemize}

These results can be used in expressions to indicate how well the suggested items are used \cite{Ricci_Rokach_Shapira-2022}:

\begin{description}
	\item[Precision] is used to indicate the fraction of recommended items that is actually relevant to a user.
	\begin{equation} \label{eqn:precision}
	\text{Precision} = \frac{ \text{True-positive items} }{ \text{Total recommended items} }
	\end{equation}
	
	\item[Recall] is used to indicate the fraction of recommended items that is actually used by a user.
	\begin{equation} \label{eqn:recall}
	\text{Recall} = \frac{ \text{True-positive items} }{ \text{Total used items} }
	\end{equation}	
\end{description}

These usage prediction measures are listed by y Agarwal and Chen (2016) \cite{Agarwal_Deepak-2016} as global ranking metrics. They also list local ranking metrics:
%% Also looked at https://towardsdatascience.com/mean-average-precision-at-k-map-k-clearly-explained-538d8e032d2
\begin{description}
	\item[Precision@K] is when only the top-K recommended items for a user in the system are included in the precision measure.
	\begin{equation} \label{eqn:precision-at-k}
	\text{Precision@}K = \frac{\text{True-positive items in top-}K \text{ recommendations}} {\text{Total items in top-}K \text{ recommendations}}
	\end{equation}
	\item[Recall@K] is when only the top-K recommended items for a user in the system are included in the recall measure.
	\begin{equation} \label{eqn:recall-at-k}
	\text{Recall@}K = \frac{\text{True-positive items in top-}K \text{ recommendations}}{ \text{Total used items}}
	\end{equation}

	\item[Average precision@K] is the sum of precision@K where the item at the $k_{th}$ position is relevant, divided by the total number of relevant items (r) in the top-K recommendations.
	\begin{equation} \label{eqn:avg-precision-at-k}
	\text{AveragePrecision@}K = \frac {1} {r} \sum_{k=1}^{K} \text{Precision@}k \cdot \text{rel}(k)
	\end{equation}
	
	\item[Mean average precision (MAP)] is used to answer the question if the most relevant recommendations are the recommendations that rank highest. The previous measures focused on a single test user that was selected, MAP calculates an average for all users (M) in the data-set.
	\begin{equation} \label{eqn:mean-average-precision}
	\text{MeanAveragePrecision@}K = \frac {1} {M} \sum_{j=1}^{M} \frac {1} {r} \sum_{k=1}^{K} \text{Precision@}k \cdot \text{rel}(k)
	\end{equation}
	
	\item[Normalized discounted cumulative gain (nDCG)] To get to nDCG we need to look at some predecessors. \textbf{Cumulative Gain} attempts to express the value of a set of recommended items. To do so, it relies on a graded relevance scale (sometimes binary) for recommended items. It is the sum of the graded relevance values of all items in a recommended list at a particular rank position (p).
	%% Also looked at https://en.wikipedia.org/wiki/Discounted_cumulative_gain
	\begin{equation} \label{eqn:cumulative-gain}
	\text{CumulativeGain}_{p} = \sum_{i=1}^{p}\text{rel}(i)
	\end{equation}
	\textbf{Discounted cumulative gain} is different in that it also penalizes highly relevant items in the list of recommendations from appearing below less relevant items.
	\begin{equation} \label{eqn:discounted-cumulative-gain}
	\text{DiscountedCumulativeGain}_{p} = \sum_{i=1}^{p} \frac{\text{rel}(i)}{\text{log}_{2}(i+1)} = \text{rel}(1) + \sum_{i=2}^{p} \frac{\text{rel}(i)}{\text{log}_{2}(i+1)}
	\end{equation}
	\textbf{Normalized discounted cumulative gain (nDCG)} exists for systems where the length of the list of recommended items can vary. Normalizing is done by relying on an Ideal Discounted Cumulative Gain (IDCG) where REL($_{p}$) is the list of recommended items up to index p ordered by relevance:
	\begin{equation} \label{eqn:ideal-discounted-cumulative-gain}
	\text{IdealDiscountedCumulativeGain}_{p} = \sum_{i=1}^{|\text{REL}_{p}|} \frac{\text{rel}(i)}{\text{log}_{2}(i+1)}
	\end{equation}
	The normalized discounted cumulative gain is formulated as:
	\begin{equation} \label{eqn:normalized-discounted-cumulative-gain}
	\text{NormalizedDiscountedCumulativeGain}_{p} = \frac{\text{DiscountedCumulativeGain}_{p}}{\text{IdealDiscountedCumulativeGain}_{p}}
	\end{equation}
	
\end{description}

\end{subappendices}

\end{document}